\documentclass[%
reprint,
superscriptaddress,
 amsmath,amssymb,
prb,
]{revtex4-2}

\usepackage{float}
\usepackage{graphicx}
\usepackage{diagbox}
\usepackage{xcolor}
\usepackage{amsmath,amssymb,amsthm,amsfonts}
\usepackage{bbm}
\usepackage{bm}% bold maths
\renewcommand{\vec }{\bm}
\renewcommand{\mathbf}{\bm}
\usepackage[normalem]{ulem}
\begin{document}

\title{Symmetry transformation of nonlinear optical current of tilted Weyl nodes \\and application to ferromagnetic $\rm MnBi_{2}Te_{4}$}

\author{Zhuocheng Lu}
\affiliation{International Center for Quantum Materials, School of Physics, Peking University, Beijing 100871, China}

\author{Ji Feng}
\affiliation{International Center for Quantum Materials, School of Physics, Peking University, Beijing 100871, China}
\affiliation{Hefei National Laboratory, Hefei 230088, China}

\begin{abstract}
A Weyl node is characterized by its chirality and tilt. We develop a theory of how $n$th-order nonlinear optical conductivity behaves under transformations of anisotropic tensor and tilt, which clarify how chirality-dependent and -independent parts of optical conductivity transform under the reversal of tilt and chirality. Built on this theory, we propose ferromagnetic $\rm MnBi_{2}Te_{4}$ as a magnetoelectrically regulated, terahertz optical device, by magnetoelectrically switching the chirality-dependent and -independent dc photocurrents. These results are useful for creating nonlinear optical devices based on topological Weyl semimetals.  
\end{abstract}

\maketitle

\section{Introduction}

The gapless excitations make Weyl semimetals an ideal platform to perform for low-energy photon detection, particularly in the infrared and terahertz regimes.\cite{liu2020} Various nonlinear optical responses, such as harmonic generations,\cite{zhang2019b,lv2021,takasan2021,tilmann2022} bulk photovoltaic effect\cite{dejuan2017,chan2017,ahn2020,yang2018b,ma2019,wang2020} and four-wave mixing\cite{gao2020,cheng2020}, have been widely studied in Weyl semimetals. Furthermore, it has recently been recognized that the geometry and topology of electronic band structures can have nontrivial impacts on optical conductivities. Nonlinear optical effects like photovoltaic effects and higher harmonic generations can be described using topological quantities involving Berry connection and curvature, which provide a unified perspective on various optical phenomena.\cite{morimoto2016b,nagaosa2017a} The circular photogalvanic effect in Weyl semimetals, for instance, is recognized as a direct measurement of the topological charge of Weyl nodes and exhibits material-independent quantization.\cite{dejuan2017} Additionally, the giant optical response of Weyl semimetals attributing to singularity enhancement by geometric quantities has been studied theoretically and experimentally.\cite{ma2019,ahn2020} As a result, Weyl semimetals have garnered substantial attention as materials of choice for exploring innovative optoelectronic devices.

Weyl nodes show up in pairs due to the no-go theorem \cite{nielsen}, and there can be as many as 160 pairs in a Weyl semimetal.\cite{chang2018c} Oftentimes, the Weyl nodes in one material can be interrelated by discrete symmetries, including rotations (proper or improper) and time-reversal, which furnishes links between optical responses between symmetry-related Weyl nodes. On the other hand, when analyzing the optical response for a single Weyl node, the symmetry of the system (with all the nodes) is unspecified. It might be straightforward to work out the relation between the optical conductivities of a pair of nodes related by rotations. However, things become obscure when it comes to symmetry operations involving time-reversal, and the difficulty arises from dissipation.\cite{ahn2020} Therefore, such symmetry analysis requires knowledge of a microscopic expression of optical conductivities. For nonlinear optical conductivities, the microscopic theory can be fairly complicated and involve multiple terms that transform differently under time-reversal. A theory to link the optical conductivities related by discrete symmetries, including rotation and time-reversal, is evidently needed.

A Weyl node in Weyl semimetals is characterized primarily by its chirality and tilt.\cite{wan2011,soluyanov2015b} In relation to optical responses, the chirality of a Weyl node determines the geometric quantities (matrix elements) that enter into the photocurrent, whereas the tilt of the spectrum determines the phase space of the electron-hole response. For instance, it has been demonstrated that finite tilt is important to generate a giant photocurrent.\cite{chan2017} And the bulk photovoltaic effect shows different behavior in low frequencies for type I and type II Weyl semimetals.\cite{yang2018b} Although it is believed that the chirality and tilt of a Weyl node have significant impacts on the direction and magnitude of photocurrent, how these basic characteristics of symmetry-related Weyl nodes reveal the relations of their nonlinear optical conductivities remains to be systematically understood. These types of relations have been largely overlooked so far because they are generally inaccessible from experiments since Weyl nodes always appear in pairs and multiple Weyl nodes usually contribute to the total optical response. However, the optical response of a single Weyl node could be interesting when the degeneracy of Weyl pairs is lifted. For example, the optical response of Weyl semimetals could come from only a single Weyl node with nonzero chirality imbalance\cite{nielsen1983,fukushima2008}. Therefore, precise statements of these types of relations are useful for further exploration of the potential optical devices based on the nonlinear optical response of Weyl nodes.

In this work, we systematically investigate how the tilt and chirality of Weyl nodes together affect the sign and magnitude of the nonlinear photocurrent. In section II, we discuss how to relate the transformation $W$ connecting the effective Hamiltonians of two Weyl nodes to their optical conductivity. We propose that a general $n$th-order optical conductivity of Weyl nodes can be decomposed as a chirality-independent part $\sigma_{0}$ and a chirality-dependent part $\sigma_{\chi}$ according to space-time inversion $P\Theta$. We also discuss their distinct geometric interpretations and show a sign-change rule of $\sigma_{0}$ and $\sigma_{\chi}$ when reversing the tilt in a certain direction. We further demonstrate the contrasting behaviors of $\sigma_{0}$ and $\sigma_{\chi}$ for odd (or even) orders when reversing the tilt or chirality, which can be generalized to multi-Weyl nodes. In section III, based on our theory, we propose ferromagnetic $\rm MnBi_{2}Te_{4}$ as a magnetoelectrically regulated optical device that carries two components of photocurrent $j_{0}$ and $j_{\chi}$, the directions of which are controlled by external electromagnetic fields. Valid parameters for experimental measurements are also discussed based on the effective Hamiltonian of Weyl node with Landau levels and length-gauge theory of nonlinear optical response.

\section{Optical conductivities of a general Weyl node}

Since a generic Weyl node is characterized solely by its chirality and tilt, it would be desirable to understand how these quantities alone, without having to consult the overall symmetry, impact the photocurrent. Although symmetry analysis is very useful in singling out zero elements in the optical conductivity tensors, this is no longer the case if we focus on the optical response of a single Weyl node. With a single Weyl node, the full symmetry of the system is unspecified, and there could be emerging symmetry at low energies. The low-energy effective Hamiltonian has the form\cite{wan2011}:
\begin{equation}
\mathcal{H}(\vec k;\vec t, A)= t_{a} k_{a} + k_{a} A_{ab} \tau_{b},
\label{eq:b1}
\end{equation}
where indices $a,b=x,y,z$ (summation implied when repeated), $\vec k$ is the wavevector, and $\tau_{a}$'s are the Pauli matrices. The non-singular matrix $A_{ab}$ is referred to as the anisotropic tensor, and correspondingly, $\chi=\text{det}A/|\text{det}A|=\pm1$ is the chirality of the Weyl node. The energy spectrum is $\varepsilon_{s}(\boldsymbol{k})= k_a t_a +s |k_a A_{ab}|$, where $s=\pm$ corresponds to conduction and valence bands, respectively. The vector $\bm t$ tilts the Weyl node, determining the shape of the Fermi surface. If $ |t_a k_a|</>|k_a A_{ab}|$ when $\boldsymbol{k}$ is along the tilt direction, the system is a type-I/type-II Weyl semimetal. A type-I Weyl semimetal has a closed Fermi surface, whereas a type-II Weyl semimetal has a Fermi surface comprised of electron and hole pockets.\cite{soluyanov2015b} Under the symmetry transformation, $\vec k$ operates as a polar c-vector, while $\vec \tau$ functions as an axial c-vector. In this context, i-/c-vectors are time-reversal symmetric/antisymmetric respectively. Consequently, Weyl nodes associated by inversion manifest opposite tilt and chirality. On the other hand, Weyl nodes linked by time-reversal display opposite tilt while preserving the same chirality. Analogous analyses can be applied to the cases of Weyl nodes associated by other symmetries. Below, we will reveal how $\bm t$ and $A$ impact the general $n$th-order photocurrent from a Weyl node described by Eq. (\ref{eq:b1}).

The photocurrent arises from the incident light's electric field, which is a superposition of multiple frequency components $E_{a}(t)=E^{\gamma}_{a}e^{-{\rm i}\omega_{\gamma}t}$, where the frequency component $\gamma$ is summed over implicitly when repeated. Here, $E^{\gamma}_{a}$ is independent of position under the long-wavelength limit. For a Weyl node described by Eq. (\ref{eq:b1}), the total $n$th-order photocurrent $j_{a}$ with a frequency $\omega_{n}$ can be expressed as 
\begin{equation}
    j_{a}^{(n)}(\omega_n)=\sum_{C(\omega_n)}\sigma^{ab_{1}...b_{n}}(\omega_{n};\omega_{\gamma_1} \cdots \omega_{\gamma_n})
    E^{\gamma_1}_{b_{1}} \cdots E^{\gamma_n}_{b_{n}},
    \label{eq:b2}
\end{equation}
where $\omega_{n}=\omega_{\gamma_1}+\cdots\omega_{\gamma_n}$, and $\sigma^{ab_{1}...b_{n}}(\cdots)$ is the $n$th-order conductivity tensor. The set $C(\omega_n)$ collects all possible selections of $n$ frequencies that sum to $\omega_{n}$. Based on the length-gauge theory of nonlinear optical response,\cite{aversa1995,sipe2000} the $n$th-order optical conductivity in Eq. (\ref{eq:b2}) can be generally written as 
\begin{equation} 
    \sigma^{a b_{1}...b_{n}} =  \int [d \boldsymbol{k}]   \Sigma^{a b_{1}...b_{n}}(\bm k),
    \label{eq:b3}
\end{equation}
in which we have omitted frequency variables and $[d \boldsymbol{k}]=\text d^3k/(2\pi)^3$. It is worth noting that the integrand $\Sigma$ is a gauge-invariant tensor, whose tensor component arises from the $k_b$-derivative, $\partial_{k_b}$. For concreteness, a detailed discussion of why the $n$th-order optical conductivity can be written in the form of Eq. (\ref{eq:b3}) are presented with examples in Appendix A.

For two Weyl nodes $\mathcal{H}$ and $\mathcal{H}^{\prime}$ related through the transformation $\mathcal{H}\mapsto\mathcal{H}^{\prime}: k_{a}\mapsto k_{a}W_{ab}$, we find their nonlinear optical conductivities, as described by Eq. (\ref{eq:b3}), have the following relationship
\begin{equation} 
    \sigma^{a b_{1}...b_{n}}\left(\mathcal{H}^{\prime}\right)=
    \sigma^{a^{\prime} b_1' \cdots b_n' }\left(\mathcal{H}\right) 
    \frac{W_{aa'} W_{b_1b_1'} \cdots W_{b_nb_n'}} {|\text{det}W|},
    \label{eq:b4}
\end{equation}
in which we assume the same chemical potentials. The product of $W_{bb'}$ factors arises from the transformation of derivatives $\partial_{k_b}$. The factor of $|\text{det}W|$ comes from the Jacobian of the momentum integral. If $|\text{det}W| = 1$, the transformation $W$ corresponds to a certain spatial symmetry operation. If $|\text{det}W|\neq{1}$, the transformation $W$ involves deformation. As a simple example, when $W_{bb^{\prime}}=-\delta_{bb^{\prime}}$, Eq. (\ref{eq:b4}) describes the relationship between the optical conductivities of two Weyl nodes connected by inversion $P$, which is $\sigma^{a b_{1}...b_{n}}\left(\mathcal{H}^{\prime}\right)=(-1)^{n+1}\sigma^{a b_{1}...b_{n}}\left(\mathcal{H}\right)$, where $n$ represents the order of response. It is essential to note that Eq. (\ref{eq:b4}) discusses the relationship between the optical conductivities of two Weyl nodes in the same spacetime, rather than the correspondence of the optical conductivity of a single Weyl node in two different spacetime coordinates. Eq. (\ref{eq:b4}) establishes a connection between the transformation $W$ applied to the effective Hamiltonian of Weyl node and the associated $n$th-order optical conductivity. Consequently, Eq. (\ref{eq:b4}) can serve as a tool for studying the nonlinear optical properties of Weyl nodes.

In Eq. (\ref{eq:b4}), both tilt and chirality are affected by the transformation $W$. For further isolation of the impacts of the transformations of tilt and chirality on the optical conductivity, we note that the $n$th-order optical conductivity for a Weyl node described by Eq. (\ref{eq:b1}) can be decomposed as
\begin{equation} 
    \sigma^{a b_{1}...b_{n}} = \sigma_{0}^{a b_{1}...b_{n}} + \sigma_{\chi}^{a b_{1}...b_{n}}
    \label{eq:b5}
\end{equation}
where the first term $\sigma_0$ is chirality-independent, whereas the second term  $\sigma_\chi$ reverses sign upon chirality reversal, i.e.,  $A\mapsto -A$. Furthermore, we recognize that two Weyl nodes related by space-time inversion $P\Theta$ have the same $\boldsymbol{t}$, while their $A$ differ by a sign. Therefore, we identify $\sigma_0$ as the portion that remains unchanged under $P\Theta$, while $\sigma_{\chi}$ represents the portion that changes sign under $P\Theta$. Additionally, the distinction between $\sigma_0$ and $\sigma_{\chi}$ is closely related to their different geometric interpretations. From the discussions in the Appendix A, we can see that quantity $r_{nm}^{a} r_{mn}^{b}$ ($r_{nm}^{a}$ is the interband Berry connection) is widely involved in the first-order to third-order optical conductivities. This quantity has a geometric interpretation of band-resolved quantum geometric tensor, which can be decomposed as $r_{nm}^{a} r_{mn}^{b}=g^{ab}_{nm}-\frac{i}{2}\Omega^{ab}_{nm}$.\cite{ahn2020, provost1980c} $g_{nm}^{ab}$ is the symmetric part that corresponds to the band-resolved quantum metric, while the antisymmetric part $\Omega^{ab}_{nm}$ is the band-resolved Berry curvature. Their names come from the fact that the Berry curvature and quantum metric can be recovered by summing over band $m$ $g_{n}^{ab}=\sum_{m\neq n}g_{nm}^{ab}$ and $\Omega_{n}^{ab}=\sum_{m\neq n}\Omega^{ab}_{nm}$. The quantum metric is a chirality-independent quantity since it can be interpreted as the distance between different quantum states. On the other hand, the Berry curvature reverses sign when chirality changes, since Weyl nodes with $\chi=\pm1$ correspond to the source (or sink) of Berry curvature in $k$ space. Here, we emphasize that \(\sigma_{0}\)/\(\sigma_{\chi}\) does not necessarily correspond to the band-resolved quantum metric/Berry curvature. The specific correspondence should be determined through a symmetry analysis of the microscopic conductivity expressions, which is discussed with examples in the Appendix B.

Eq. (\ref{eq:b5}) discusses the impact of chirality changes on optical conductivity. Next, we want to explore the influence of tilt variations on optical conductivity. Without loss of generality, we can always choose a coordinate system where $A$ is diagonal. In such a scenario, we discover that two Weyl nodes, possessing the same $A$ and tilts that are identical across all components except for an opposite $c$ component, are linked by the symmetry operation $\mathcal{M}_c=M_{c}P\Theta$, where $M_{c}$ is mirror reflection in $c$ direction. Aided by Eq. (\ref{eq:b4}) and Eq. (\ref{eq:b5}), we ascertain that the optical conductivities of these two Weyl nodes relate as follows:
\begin{equation}
    \begin{aligned} 
       &\sigma^{a b_{1}...b_{n}}_{0} (\mathcal{M}_c\mathcal{H}\mathcal{M}_c^{\dagger}) = (-1)^{\alpha_{c}} \sigma^{a b_{1}...b_{n}}_{0} (\mathcal{H})  \\
       &\sigma^{a b_{1}...b_{n}}_{\chi} (\mathcal{M}_c\mathcal{H}\mathcal{M}_c^{\dagger}) = (-1)^{\alpha_{c}+1} \sigma^{a b_{1}...b_{n}}_{\chi} (\mathcal{H}),
       \label{eq:b6}
    \end{aligned} 
\end{equation} 
in which $\alpha_{c}$ marks how many times $c$ appears in superscript $ab_1\cdots b_n$. In Eq. (\ref{eq:b6}), $\sigma_{0}$ and $\sigma_{\chi}$ exhibit distinct changes when $t_{c}\rightarrow-t_{c}$. This arises from the fact that chirality changes sign under $P\Theta$. Eq. (\ref{eq:b6}) can be helpful when we study the nonlinear optical conductivities of two Weyl nodes related to each other by transformation $t_c\mapsto  - t_c$ dictated by the symmetry relating the Weyl pair. In particular, for a Weyl node with $t_c=0$, $t_c\mapsto  - t_c$ will immediately reveal the selection rule for $\sigma^{a b_{1}...b_{n}}_{0}$/$\sigma^{a b_{1}...b_{n}}_{\chi}$.

\begin{table}
    \caption{Transformations of $n$th-order optical conductivities of a Weyl node under $\Theta$, $P$, and $P\Theta$. Since these transformations depend only on whether the order of response is even or odd, we use $\sigma^{\text{odd/even}}$ to denote the tensor for an odd/even-order response. The plus (minus) one below $\vec t$/$A$/$\sigma$ means the sign of tilt/chirality/conductivities is unchanged (reversed) under the corresponding symmetry operation. We observe that, $\sigma_{0}^{\rm odd}$ is insensitive to the reversal of tilt and chirality, while $\sigma_{\chi}^{\rm odd}$ changes sign due to the reversal of tilt or chirality. On the other hand, the sign of $\sigma_{0}^{\rm even}$/$\sigma_{\chi}^{\rm even}$ exclusively depends on the tilt/chirality.} 
    \label{tab:1}
    \begin{tabular}{cccccc} 
    \hline\hline  
    &$(\vec t,A)$     
    &$\sigma^{\text{odd}}_{0}$    
    &$\sigma^{\text{odd}}_{\chi}$ 
    &$\sigma^{\text{even}}_{0}$ 
    &$\sigma^{\text{even}}_{\chi}$ \\ 
    \hline
    $\Theta$    &$(-1,+1)$   &$+1$    &$-1$    &$-1$    &$+1$  \\ 
    $P$    &$(-1,-1)$   &$+1$    &$+1$    &$-1$    &$-1$  \\
    $P \Theta$  &$(+1,-1)$   &$+1$    &$-1$    &$+1$    &$-1$  \\ 
    \hline\hline
    \end{tabular}
    \end{table}

Let us comment on how the results can be useful in analyzing optical responses in Weyl semimetals, before going to concrete examples. Using Eq. (\ref{eq:b4}-\ref{eq:b6}), we can then quickly determine whether a conductivity tensor component of a Weyl node should be zero and whether it corresponds to $\sigma_{0}$ or $\sigma_{\chi}$. This leads to the knowledge that how optical conductivities of a symmetry-related Weyl pair are interrelated based solely on  $\vec t$ and $A$. In particular, this is made possible regardless of the frequency-dependence of the optical conductivity tensor and whether the Weyl nodes are type-I or type-II. These are done without requiring the knowledge of complete symmetry of the Weyl nodes or recourse to microscopic expressions of the conductivities even if the symmetry operation involves time-reversal. As a quick demonstration, we can work out how the conductivities of a Weyl node transform under time-reversal $\Theta$, inversion $P$, and the space-time inversion $P\Theta$, as shown in Table \ref{tab:1}. According to Eq. (\ref{eq:b5}), $P\Theta$ changes the sign of $\sigma_{\chi}$ while keeping $\sigma_{0}$ unchanged. On the other hand, according to Eq. (\ref{eq:b6}), $\Theta$ changes the sign of $\sigma_{\chi}^{\text{odd}}$ and $\sigma_{0}^{\text{even}}$, while keeping $\sigma_{\chi}^{\text{even}}$ and $\sigma_{0}^{\text{odd}}$ unchanged. Then, how $\sigma_{0}$ and $\sigma_{\chi}$ changes under $P$ can be directly deduced from the composite operation of $P\Theta$ and $\Theta$. It should be noted that results in Table \ref{tab:1} is also a consequence of how $\sigma_{0}$ and $\sigma_{\chi}$ transform under reversal of chirality and how the $n$th-order optical conductivity transfroms under $P$. Before ending this section, we emphasize that results in Eqs. (\ref{eq:b5}-\ref{eq:b6}) and Table \ref{tab:1} can be generalized to the multi-Weyl nodes\cite{fang2012,li2021}. The detailed discussion is presented in Appendix C.

\section{Magnetoelectric switch}   

\begin{figure}[htbp]
    \centering
    \includegraphics[width=7cm]{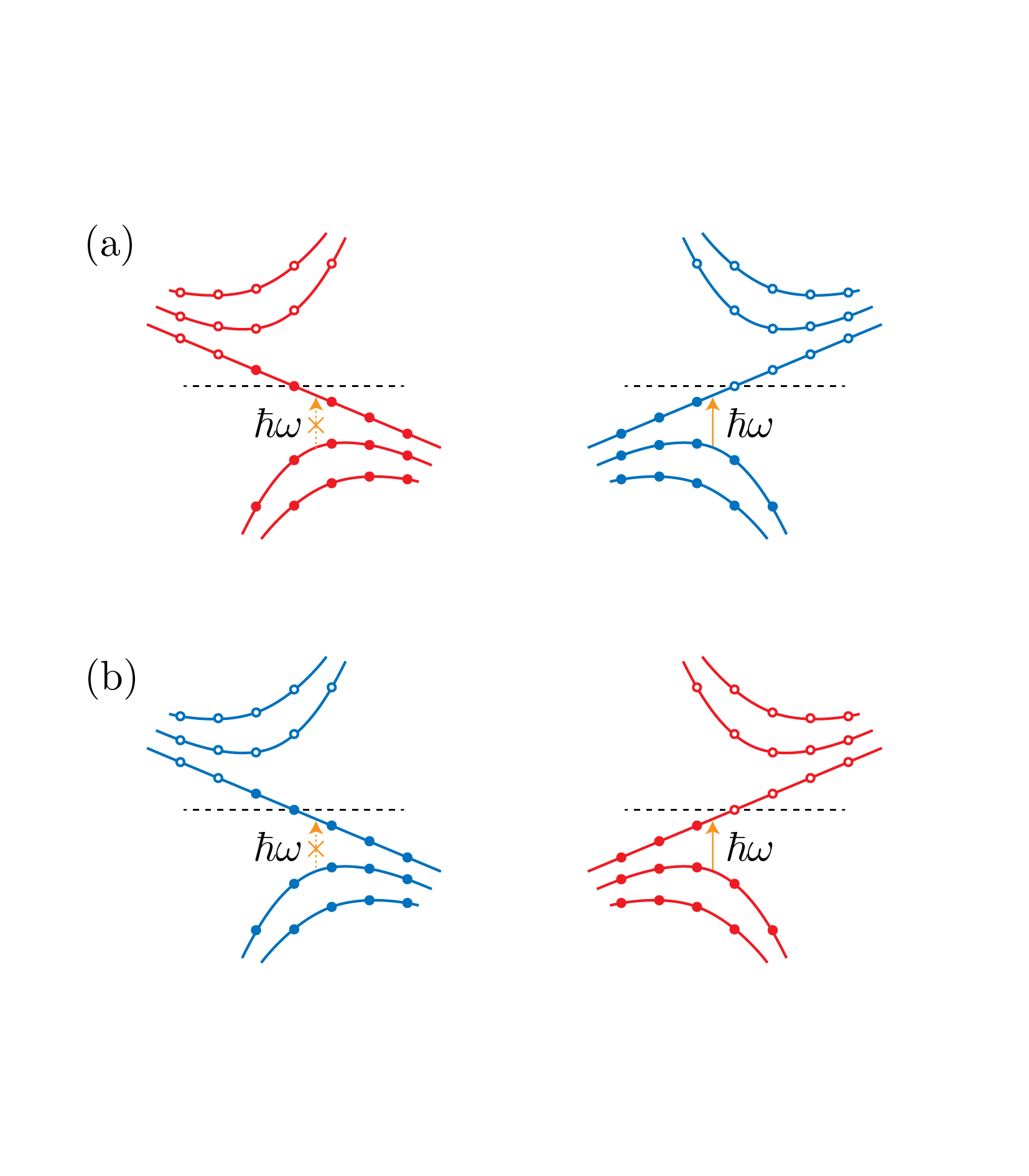}
    \caption{Schematic of the switching mechanism of the photocurrent by chirality imbalance. Nodes I and II are depicted in (a) while nodes II$'$ and I$'$ are depicted in (b). A Weyl node is colored red/blue to indicate positive/negative chirality. Solid circles indicate occupied states while hollow circles indicate unoccupied states. Owing to the chiral magnetic field, each pair of nodes is differently populated. The black dashed line indicates the Fermi level without chirality imbalance.}
    \label{fig:fig01}
\end{figure}

The foregoing analysis highlights the possibility of controlling the total photocurrent of the system by switching the Weyl nodes with different combinations of tilt and chirality. In this section, we propose a mechansim for magnetoelectrically switching using Weyl semimetals, making use of the formulas we developed in the last section. We consider a magnetic Weyl semimetal with a pair of Weyl nodes, dubbed I and II, interelated by inversion. Suppose the material is a soft magnet whose magnetization can be easily trained by an external magnetic field $\bm B$. Without loss of generality, node I tilts along positive $z$-direction and with a chirality $\chi=+1$, and correspondingly node II tilts in negative $z$-direction with $\chi=-1$. Upon magnetization reversal by reversing the $B$-field, node I/II becomes its time-reversal counterparts I$'$/II$'$, which has opposite tilt and the same chirality. We also introduce an external dc electric field $\bm E_{\rm dc}$ that is parallel (or antiparallel) to $\bm B$, which together with the $B$-field can induce a chemical potential difference within a pair of Weyl nodes at steady states, 
\begin{equation}
\begin{aligned} 
    \delta \mu = \mu_{\text{I/I}'} - \mu_{\text{II/II}'}\propto \bm E_{\text{dc}}\cdot \bm B,
    \label{eq:c1}
\end{aligned}
\end{equation}
owing to the chirality imbalance originating from chiral anomaly.\cite{nielsen1983,fukushima2008} Then for a given incident photon energy $\hbar\omega$, we can combine the chirality imbalance with the Pauli blocking to selectively activate Weyl nodes for photocurrent. As schematically depicted in Fig.\ref{fig:fig01}, the vertical electron-hole excitations are allowed (solid orange arrows)/forbidden (dashed light orange arrows) for activated/deactivated Weyl node when an appropriate $\hbar\omega$ is considered. Therefore, chirality imbalance can work as a switch of optical response of nodes I,II,I$'$,II$'$ by reversing the directions of $B$-field and $E_{\rm dc}$-field, which makes it possible for us to control the sign of tilt and chirality of Weyl node for photocurrent. In the setup depicted, I or I$'$ is activated for $\delta \mu<0$, and II or II$'$ is activated for $\delta \mu>0$. An important observation from Table \ref{tab:1} is that $\sigma_{0}^{\text{even}}$ transforms as $\vec t$, and  $\sigma_{\chi}^{\text{even}}$ transforms as $A$. Consequently, by selectively activating Weyl node with $B$-field and $E_{\rm dc}$-field, a magnetoelectric switch can be envisioned that couples the tilt and chirality of Weyl node to the direction of its second-order photocurrent.

\begin{figure}[htbp]
    \centering
    \includegraphics[width=7cm]{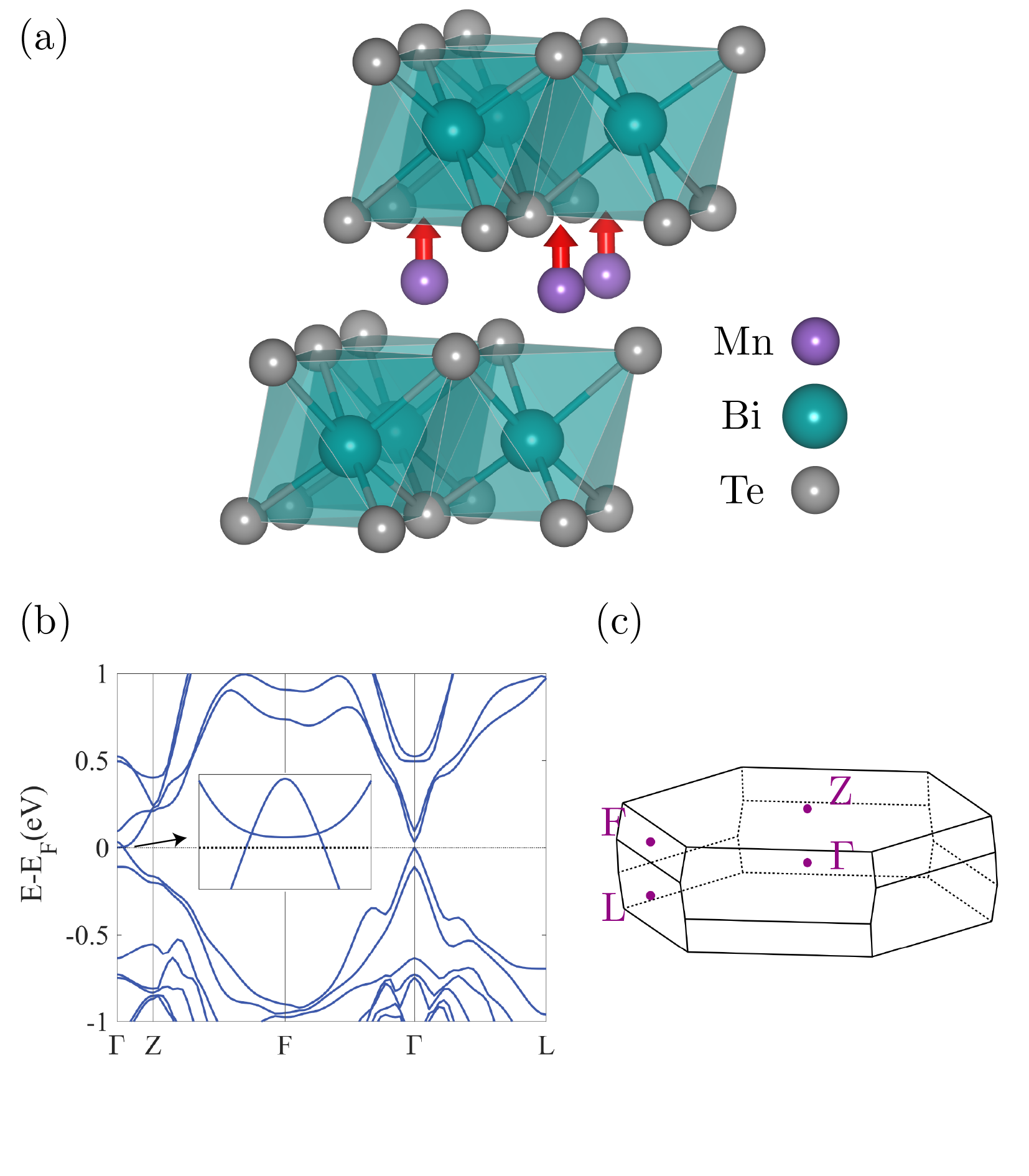}
    \caption{Crystal and magnetic structures of ferromagnetic $\rm MnBi_{2}Te_{4}$ are shown in (a), and the DFT band structure in a 2 eV window around the Fermi level is shown in (b). (a) The red arrows represent magnetic moments on Mn. (b) The DFT band structure is shown as black lines. Weyl nodes $\rm I$ and $\rm II$ on the high symmetric lines ${\rm \bar{Z}}-\Gamma-{\rm Z}$ are zoomed in and shown in the inset. (c) The high symmetry points are exhibited in the Brillouin zone with purple dots.}
    \label{fig:fig02}
\end{figure}

The magnetoelectrically switching mechanism described above may be realized in ferromagnetic $\rm MnBi_{2}Te_{4}$. Though the ground state of $\rm MnBi_{2}Te_{4}$ is an antiferromagnetic topological insulator,\cite{zhang2019,otrokov2019,li2019d} ferromagnetic $\rm MnBi_{2}Te_{4}$ obtained under high magnetic field is computationally predicted to be Weyl semimetal that has only a pair of Weyl nodes,\cite{zhang2019,li2019d} supported by preliminary experimental data.\cite{lee2021} $\rm MnBi_{2}Te_{4}$ comprises negatively charged Bi-Te layers neutralized with intercalated magnetic Mn(II) cations in hexagonal close packing, as shown in Fig.\ref{fig:fig02}(a). When the $B$-field aligns the magnetic moments in $z$-direction, the magnetic point group of ferromagnetic $\rm MnBi_{2}Te_{4}$ is $\bar{3}m^{\prime}$, and the pair of Weyl nodes, which can be marked as I and II, are bound to the high-symmetry $k_z$ axis and tilt oppositely in the $z$-direction. Reversal of the $B$-field brings about Weyl nodes I$'$ and II$'$, which are the time-reversal images of I and II, respectively. The tilt directions and chiralities of I,II and their time-reversal images are the same as those previously proposed. We will consider ferromagnetic $\rm MnBi_{2}Te_{4}$ magnetized along  $\pm z$-direction only. 

The electronic structure of ferromagnetic $\rm MnBi_{2}Te_{4}$ is shown in Fig.\ref{fig:fig02} (b). Weyl nodes $\rm I$ and $\rm II$ are on the high symmetric lines $\bar{{\rm Z}}-\Gamma-{\rm Z}$. When reversing the direction of the $B$-field, the spectrum stays unchanged while chirality of Weyl nodes is switched. First-principles calculations were performed using Vienna ab initio Simulation Package equipped with the projector-augmented-wave potentials \cite{kresse1996}.  For structural relaxation, the exchange-correlation interactions were considered in the generalized gradient approximation (GGA) with Perdew-Burke-Ernzerhof scheme \cite{perdew1996}.  For self-consistent calculation and band structure calculation, we consider modified Becke-Johnson  method. The correlation of $3d$ orbitals of $\text{Mn}$ is partially treated with GGA+$U$ formalism, with an isotropic $U =4.0$ eV. The spin-orbital coupling is included in self-consistant calculations. A 350 eV cutoff energy and $6\times6\times6$ $\bm k$-point sampling is considered.

Since ferromagnetic phase of $\rm MnBi_{2}Te_{4}$ requires at least a 7.7 T $B$-field,\cite{lee2019} the effect of Landau levels should be considered. We fit the parameters of the low-energy effective Hamiltonian of Weyl nodes with the DFT band structure of ferromagnetic $\rm MnBi_{2}Te_{4}$. Introducing the $B$-field in $z$-direction with vector potential $\bm a = Bx\hat y/\hbar$, the Weyl nodes then can be described by 
 \begin{equation}
     \mathcal H = t v_t k_z + \chi[v_{\parallel}k_z\sigma_z + v_{\perp}(k_x \sigma_x + (k_y+ea_y) \sigma_y)].
     \label{eq:c3}
 \end{equation}
 Fermi velocities in the plane are same because of threefold rotational symmetry along $z$-direction. $v_{t}={\rm 0.494 eV \cdot}$\AA, $v_{\parallel}={\rm 0.625 eV \cdot}$\AA \ and $v_{\perp}={\rm 1.975 eV \cdot}$\AA. $t=+1$ for nodes I, II$'$ while $t=-1$ for nodes II, I$'$. $\chi= +1$ for nodes I, I$'$ while $\chi= -1$ for nodes II, II$'$. Here, the electron charge is $-e ({\rm i.e.},e>0)$. Since the two Weyl nodes are well-separated in momentum space, the relaxation time of chirality-violating scattering $\tau_{\rm cv}$ is much larger than that of chirality-preserving scattering $\tau_{\rm cp}$. At steady states, the chiral anomaly can be treated as the shift of chemical potential $\chi\delta\mu/2$ for the two Weyl nodes.  The chemical potential without chirality imbalance is $\mu_{0}=-$6meV. The energy of the $n$th ($n$ an integer) Landau level at $\vec k =[k_y,k_z]$ is 
\begin{equation}
\varepsilon_{n\vec k} = t v_t k_z +\chi \operatorname{sgn} (n)\eta_n,
\label{eq:c4}
\end{equation}
in which $\eta_n = \sqrt{2|n| \hbar^2 v_\perp^2/\ell^2+\eta_{\parallel}^2}\ (n\neq0)$, $\ell = \sqrt{\hbar/eB}$ is the magnetic length, $\eta_{0}=\eta_{\parallel}= v_{\parallel}k_z$, and the sign function $\operatorname{sgn}(n)$ equals $-1$ if $n=0$. The details of calculations of eigenfunctions and velocity matrix elements are presented in Appendix D, from which we can see that the optical transitions are nonzero only between Landau levels that satisfy $|m| - |n|=\pm 1$.

Now, we are in a position to verify our proposal of magnetoelectrically regulated photocurrent in ferromagnetic $\rm MnBi_{2}Te_{4}$. For a straightfoward dc measurement, we consider that the system is illuminated by light propagating in $z$-direction. The optical electric field can be written as $E(t)=|E|(\cos \omega t, \eta \sin \omega t, 0)$. The incident light is linearly polarized when $\eta=0$, or circular if $\eta=\pm 1$, or elliptical somewhere in between. The optical electric field $E(t)$ has no contributions to the chiral anomaly since it is perpendicular to the $B$-field. The total dc current in $z$-direction has two components $j_z^{\rm stat}$ and $j_z^{\rm opt}$. $j_z^{\rm stat}$ depends only on the $E_{\text{dc}}$-field while $j_z^{\rm opt}$ arises from the illumination of light. $j_z^{\rm stat}$ can be taken as background since it is insensitive to the change of light. Then, the dominant dc current flows in $z$-direction comes from the photovoltaic effect of the activated Weyl nodes, that is
\begin{equation}
\begin{aligned} 
    j_{z}^{\rm (2)}(\eta)= j_{0} + \eta j_{\chi}
    \label{eq:c2}
\end{aligned} 
\end{equation} 
with $j_{0} \equiv \tilde{\sigma}_{0}|E|^{2} = (\sigma^{zxx}_{0}+\sigma^{zyy}_{0})|E|^{2} $ and $j_{\chi} \equiv \tilde{\sigma}_{\chi}|E|^{2} = {\rm i}(\sigma^{zyx}_{\chi}-\sigma^{zxy}_{\chi})|E|^{2}$. Other optical conductivity components are zero for Weyl nodes tilting in $z$-direction and $A$ is diagonal according to Eq. (\ref{eq:b6}). Therefore, the validation of magnetoelectrically regulated photocurrent in ferromagnetic $\rm MnBi_{2}Te_{4}$ is verified, since direction of the photocurrent $j_{0}$/$j_{\chi}$ is determined exclusively by the tilt/chirality, and $j_{0}$ and $j_{\chi}$ can be distinguished by varying $\eta$. The proposed magnetoelectric switch based on ferromagnetic $\rm MnBi_{2}Te_{4}$ is an illustration of analyzing the photocurrent using our result introduced in section II, without the knowledge of the complete symmetry of the Weyl nodes. 

Next, we would like to discuss the experimental parameters of magnetoelectrically switching by computing the second-order dc photocurrent of Weyl nodes in ferromagnetic $\rm MnBi_{2}Te_{4}$. Based on the length-gauge theory of nonlinear optical response, dominant contributions of the photogalvanic effect are from the injection current and shift current in the clean limit.\cite{aversa1995,sipe2000} Upon comparing the results presented in Table \ref{tab:1} with the symmetry analysis of injection current and shift current,\cite{ahn2020} we observe that the circular shift current and linear injection current correspond to $\sigma_{0}$, while linear shift current and circular injection current correspond to $\sigma_{\chi}$. Further, referring to Eq. (\ref{eq:c2}), we note that only the linear injection current contributes to $j_{0}$, while only the circular injection current contributes to $j_{\chi}$. For the Weyl nodes described by Eq. (\ref{eq:c3}), injection current can be written as
\begin{equation}
\begin{aligned}
\sigma^{abc} = -\frac{\tau_{\rm cp} e^{3}}{4\pi^2\hbar^{2}\ell^2}\int dk_z \sum _{n,m} f_{nm} \Delta _{mn}^{a} r_{nm}^{c} r_{mn}^{b} \delta ( \omega _{mn} -\omega ) 
\label{eq:c5}
\end{aligned} 
\end{equation}
where $\tau_{\rm cp}$ is the relaxation time of chirality-preserving scattering, $f_{nm}=f_{n}-f_{m}$ with $f_{n}=f(\varepsilon_n)$ is the occupation number of the $n$th band, $\Delta_{mn}^{a}=v_{m}^{a}-v_{n}^{a}$ with $v_{n}^{a}$ is the group velocity of $n$ band, $r^{c}_{nm}$ is the interband Berry connection. Given that the linear (circular) injection current is the real (imaginary) part of the conductivity, we have $\sigma_{0}^{abc} = {\rm Re}\{\sigma^{abc}\}$ and $\sigma_{\chi}^{abc} = {\rm Im}\{\sigma^{abc}\}$. Then $\tilde{\sigma}_{0}$ and $\tilde{\sigma}_{\chi}$ can be accessed by inserting Eq.(\ref{eq:c5}) back to their definitions in Eq.(\ref{eq:c2}). 

\begin{figure}[htbp]
    \centering
    \includegraphics[width=7cm]{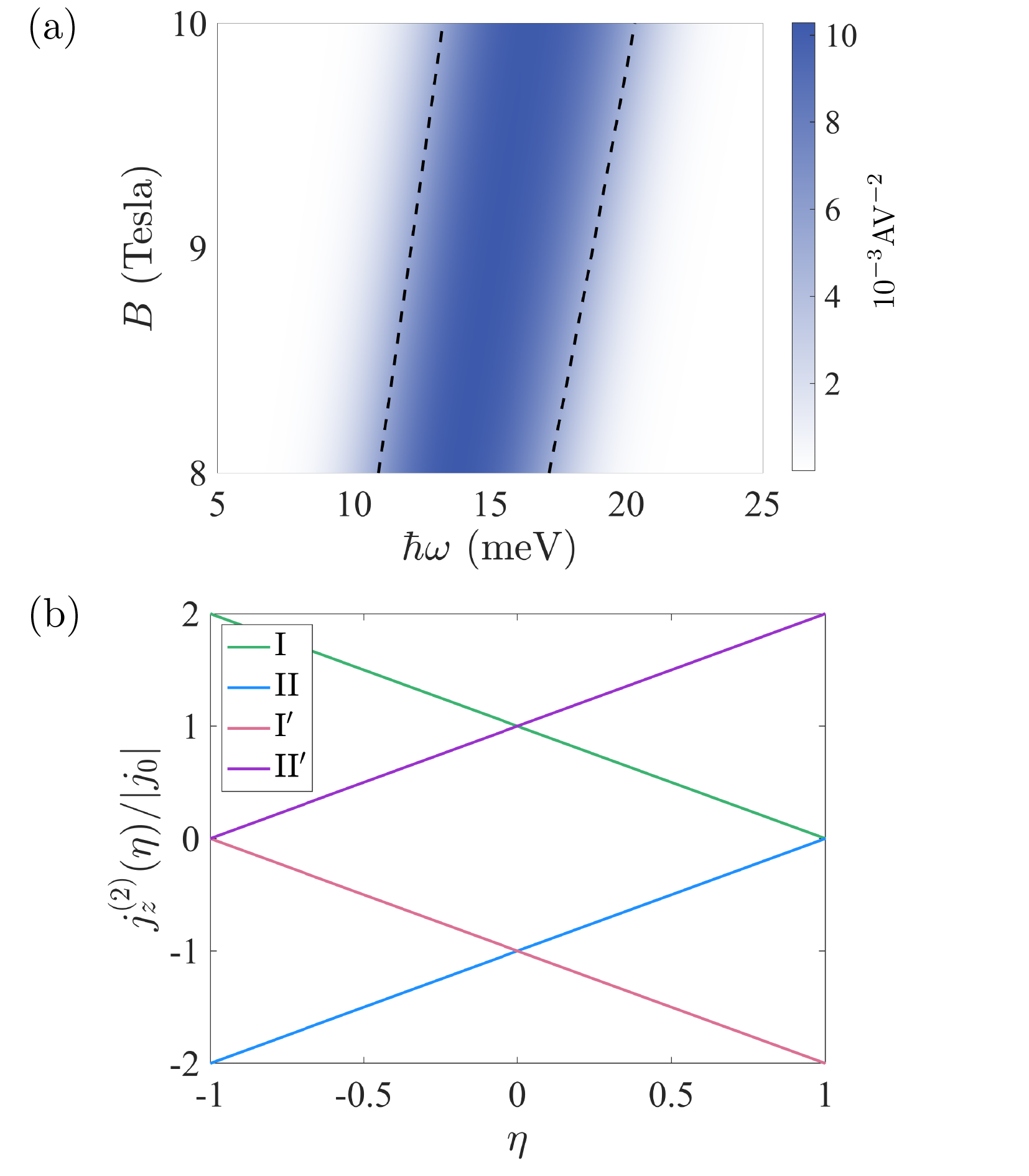}
    \caption{Numerical calculations of conductivities $\tilde{\sigma}_{0}$ and $\tilde{\sigma}_{\chi}$ of nodes I, II, I$'$, II$'$. (a) The magnitude of $\tilde{\sigma}_{0}$ and $\tilde{\sigma}_{\chi}$, which only differ in signs for all nodes: 
    $\tilde{\sigma}_{0}(\text{I}) =\tilde{\sigma}_{0}(\text{II}')=\tilde{\sigma}_{\chi}(\text{II})=\tilde{\sigma}_{\chi}(\text{II}') $
    $=-\tilde{\sigma}_{0}(\text{I}') =-\tilde{\sigma}_{0}(\text{II})=-\tilde{\sigma}_{\chi}(\text{I})=-\tilde{\sigma}_{\chi}(\text{I}')$. Dashed lines indicate the upper and lower frequency boundaries outside of which the photocurrents of two nodes cancel out. (b) Total second-order dc photocurrent $j_{z}^{(2)}(\eta)$ of nodes I,II,I$'$,II$'$ divided by $|j_{0}|$ as a function of $\eta$.}
    \label{fig:fig03}
\end{figure}

\begin{figure}[htbp]
    \centering
    \includegraphics[width=7cm]{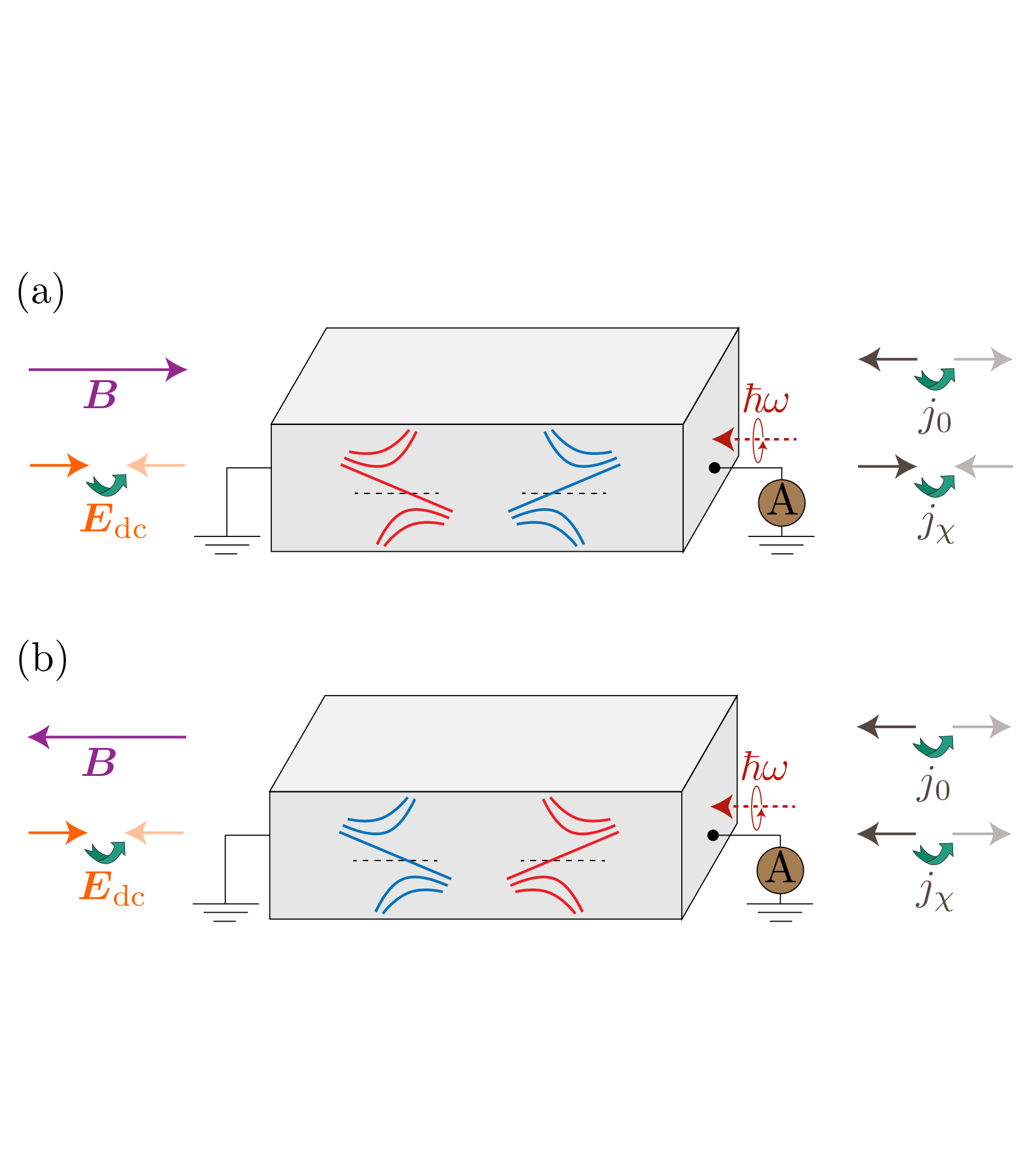}
    \caption{Magnetoelectrically regulated photocurrent in ferromagnetic $\rm MnBi_{2}Te_{4}$. (a) Configuration with $B$-field in positive $z$-direction. (b) Configuration with $B$-field in negative $z$-direction. The directions of $B$-field, $E_{\rm dc}$-field, and resulting $j_{0}$, $j_{\chi}$ are exhibited. Spectra of Weyl nodes with Landau levels are schematically shown in the front of cubic samples. Zeroth Landau level is the straight line that crosses the Fermi level. The black dashed line denotes the Fermi level. Landau levels that are below zeroth Landau level are all immersed in Fermi sea.}
\label{fig:fig04}
\end{figure}

The second-order dc optical conductivities $\tilde{\sigma}_{0}$ and $\tilde{\sigma}_{\chi}$ computed for ferromagnetic $\rm MnBi_{2}Te_{4}$ as a function of $\hbar\omega$ and $B$ are shown in Fig.\ref{fig:fig03} (a). It is easy to show that $\tilde{\sigma}_{0}$ and $\tilde{\sigma}_{\chi}$ of nodes I,II,I$'$,II$'$ all have the same magnitudes apart from different signs, so we only plot $\tilde{\sigma}_{0}$ of node I in Fig.\ref{fig:fig03} (a). The two black dashed lines represent the critical values for optical excitations of the two Weyl nodes with different chemical potentials, respectively. For Weyl node described by Eq. (\ref{eq:c3}), only valid optical excitations are from Landau level $n=\mp 1$ (for $\chi=\pm 1$) to zeroth Landau level. Then the critical value for nonzero optical excitations $\hbar\omega_{\rm c}$ is given by
\begin{equation}
\hbar \omega_{\rm c} = \sqrt{2\hbar v_{\perp }^{2} eB+\mu^{2}\frac{v_{\| }^{2}}{( v_{\| } -v_{t})^{2}}} + \mu\frac{v_{\| }}{v_{\| } -v_{t}},  
\label{eq:c6}
\end{equation}
which is at the $k_z$ where zeroth Landau level crosses the Fermi level. We can see that $\hbar \omega_{\rm c}$ is smaller for Weyl node with Fermi level $\mu$ lowered by chirality imbalance within a Weyl pair. When the $E_{\rm dc}$-field is parallel to the $B$-field, $\delta\mu>0$, and the dashed black lines on the left and right correspond to the Weyl nodes with $\chi=\mp 1$. Conversely, when the $E_{\rm dc}$-field is antiparallel to the $B$-field, $\delta\mu<0$, and the dashed black lines on the left and right correspond to the Weyl nodes with $\chi=\pm 1$. Also, it should be noted that this critical value is the same for nodes I,II,I$'$,II$'$ with same magnitude of Fermi levels $\mu$ and $B$-field.

For the range of energy of light showing the switching effect to be not too small, $\delta\mu$ should be comparable to $\mu_{0}$. Since $\mu_{0}=-6$ meV, here we take $\delta\mu=4$ meV, along with a $B$-field in the range from 8T to 10T. The Landau level spacing sets roughly the minimum energy of light (about 10 meV). And we plot $\hbar\omega$ between 5 and 25 meV in Fig. \ref{fig:fig03} (a). For a given $B$-field, when the energy of light is too small to excite either one of the Weyl nodes, both $\tilde{\sigma}_{0}$ and $\tilde{\sigma}_{\chi}$ are zero. When the light energy increases and meets the left black dashed line, only one Weyl node is activated for optical response while the other is Pauli blocked due to the chiral imbalance. In this case, $\tilde{\sigma}_{0}$ and $\tilde{\sigma}_{\chi}$ acquire nonzero values, the signs of which are determined by the tilt and chirality of the activated node, respectively. As light energy increases further and exceeds the right black dashed line, $\tilde{\sigma}_{0}$ and $\tilde{\sigma}_{\chi}$ are zero again since both Weyl nodes are activated. Therefore, the colored areas in Fig.\ref{fig:fig03} (a) correspond to optical response from a single activated Weyl node. Nonzero $\tilde{\sigma}_{0}$ and $\tilde{\sigma}_{\chi}$ outside the two black dashed lines originate from the small but finite relaxation time. Fig.\ref{fig:fig03} (b) shows how the total second-order dc photocurrents $j_{z}^{(2)}(\eta)$ vary with $\eta$, which demonstrates the possibility of controlling $j_{z}^{(2)}(\eta)$ and identifying Weyl nodes by varying the polarization of light, as have been discussed after Eq. (\ref{eq:c2}). When $\eta=0$, the only nonzero contribution to $j_{z}^{(2)}(\eta)$ comes from $j_{0}$. The slope of each line corresponds to the sign of $j_{\chi}$. Moreover, $j_{z}^{(2)}(\eta)$ for nodes I and II (or I$'$ and II$'$) sum up to zero at any given $\eta$ due to the inversion symmetry of the system.

Based on the discussion above, we propose ferromagnetic $\rm MnBi_{2}Te_{4}$ as a magnetoelectrically regulated terahertz optical device, as illustrated in Fig.\ref{fig:fig04}. In Fig.\ref{fig:fig04}, the upper and lower panels correspond to two different magnetic configurations, namely, $B$-field along the positive $z$-direction (upper panel) and along the negative $z$-direction (lower panel). The direction of the $B$-field is indicated on the left side of the figure by purple arrows. Under a given magnetic configuration, inverting the $E_{\rm dc}$-field leads to the reversal of the activated Weyl node. The $E_{\rm dc}$-field direction, along with the directions of $j_{0}$ and $j_{\chi}$, is depicted on the left and right sides of the device, respectively. Let's take the case of the $B$-field along the positive $z$-direction as an example to illustate how the device works. When the $E_{\rm dc}$-field is oriented along the positive $z$-direction (orange arrow), $\delta\mu>0$, causing the Weyl node with $\chi=-1$ to have lower chemical potential. Consequently, it can be activated by smaller $\hbar\omega$, while the Weyl node with $\chi=+1$ is deactivated due to Pauli blocking. The dark gray arrows on the right side of the figure illustrate the directions of $j_{0}$ and $j_{\chi}$. When the direction of $E_{\rm dc}$-field is reversed to the negative $z$-direction (pale orange arrow), the Weyl node with $\chi=+1$ is activated, while the Weyl node with $\chi=-1$ is deactivated. The directions of $j_{0}$ and $j_{\chi}$ are also reversed to the direction of the light gray arrows.

\section{Conclusion}
In summary, our study presented a systematic approach to characterize the nonlinear optical conductivity of Weyl nodes based on their tilt and chirality. We have explored how the transformations of the tilt and chirality of the Weyl node impact its nonlinear optical conductivity. We have shown that the general $n$th-order optical conductivity of Weyl nodes can be decomposed into a chirality-independent part $\sigma_{0}$ and a chirality-dependent part $\sigma_{\chi}$. Furthermore, we have discussed how $\sigma_{0}$ and $\sigma_{\chi}$ respond to the change of tilt and chirality. Our approach can be used to characterize the optical conductivities of Weyl nodes and to analyze the interrelation of optical conductivities between a pair of symmetry-related Weyl nodes. Remarkably, our theory does not require a comprehensive understanding of the complete symmetry or microscopic expressions of the optical conductivities. To showcase the utility of our approach, we propose ferromagnetic $\rm MnBi_{2}Te_{4}$ as a magnetoelectrically regulated optical device that can generate two independent photocurrent components, $j_{0}$ and $j_{\chi}$. Their signs can be cooperatively controlled by the directions of the $B$-field and $E_{\rm dc}$-field.

\begin{acknowledgments}
We acknowledge the financial support from 
the National Key R\&D Program of China (Grants No.
2018YFA0305601 and No. 2021YFA1400100), the National Natural Science Foundation of China
(Grants No. 12274003, No. 11725415, and No. 11934001), and the Innovation Program for Quantum Science and Technology (Grant
No. 2021ZD0302600).
\end{acknowledgments}

\section{Appendix}
\subsection{Brief introduction to length-gauge theory}
\setcounter{equation}{0}
\renewcommand\theequation{A\arabic{equation}} 
In this section, we will briefly review the length-gauge theory of nonlinear optical response. And we will make examples to show why the general nonlinear optical conductivity can be expressed in the form of Eq. (\ref{eq:b3}).

The perturbation of electric field of light can be handled either in velocity-gauge or length-gauge, which are two equivalent treatments up to an unitary transformation. In the velocity-gauge, the Hamiltonian is obtained with $\boldsymbol{p} {\rightarrow} \boldsymbol{p}+e\boldsymbol{A}$. However, when it comes to practical calculations, velocity-gauge treatment suffers from several troublesome drawbacks.\cite{ventura2017} On the contrast, the length-gauge theory developed by Sipe and others\cite{aversa1995,sipe2000} are free from these drawbacks and widely implemented in the study of nonlinear optical response. In the length-gauge, the system is perturbed by the electric dipole
\begin{equation}
    \begin{aligned}  
H=H_{0}+H_{\text{scat}}+e\boldsymbol{r}\cdot\boldsymbol{E}(t).
\label{eq:aa1}
\end{aligned}
\end{equation}
$H_{0}$ is the unperturbed Hamiltonian, $H_{\text{scat}}$ represents all the scatttering effects. The electric field can be written as $\boldsymbol{E}(t)=\boldsymbol{E}^{\gamma}e^{-{\rm i}\omega_{\gamma}t}$ in the long wavelength limit. Index $\gamma$ indicates the summation over different frequency component of light and complex conjugate of certain frequency component. For example, the electric field of monochromatic light is described by $\boldsymbol{E}(t)=\boldsymbol{E}(\omega)e^{-{\rm i}{\omega}t}+\boldsymbol{E}(-\omega) e^{{\rm i}{\omega}t}$, where $\boldsymbol{E}^*(\omega)=\boldsymbol{E}(-\omega)$.

The vital problem of length-gauge theory lies in how to treat the position operator in the Bloch basis. The matrix elements of $\boldsymbol{r}$ are easier to handle by distinguishing its intraband part and interband part by $\boldsymbol{r}=\boldsymbol{r}_{i}+\boldsymbol{r}_{e}$. The matrix elements are 
\begin{equation}
\begin{aligned}  
& \langle n\boldsymbol{k}|r_{i}^{b} |m\boldsymbol{k}^{\prime}\rangle =\delta _{nm}\left[ \delta\left(\boldsymbol{k}-\boldsymbol{k}^{\prime}\right) \xi _{nn}^{b}+{\rm i}\nabla _{k_{b}} \delta \left(\boldsymbol{k}-\boldsymbol{k}^{\prime}\right)\right] \\
& \langle n\boldsymbol{k}|r_{e}^{b} |m\boldsymbol{k}^{\prime}\rangle = \delta\left(\boldsymbol{k}-\boldsymbol{k}^{\prime}\right) r _{nm}^{b}. 
\label{eq:aa2}
\end{aligned}
\end{equation} 
$|n\boldsymbol{k}\rangle $ is the Bloch state, $r_{nm}^{b}=( 1-\delta _{nm}) \xi _{nm}^{b}$, $\xi _{nm}^{b}$ is Berry connection $\xi _{nm}^{b}=\langle u_{n\boldsymbol{k}}|{\rm i}\nabla_{k_{b}} |u_{m\boldsymbol{k}}\rangle$, $|u_{m\boldsymbol{k}}\rangle$ is the periodic part of Bloch state. We can see that matrix elements of the intraband part are highly singular because a $k_b$-derivative of Dirac delta function is involved. Fortunately, intraband part $\boldsymbol{r}_{i}$ only appears in the commutators in the derivation, the matrix element of which is no longer singular, 
\begin{equation}
\begin{aligned}  
& \langle n\boldsymbol{k}|\left[ r_{i}^{b} ,S^{c} \right] |m\boldsymbol{k}^{\prime}\rangle =\delta \left( \boldsymbol{k}-\boldsymbol{k}^{\prime}\right){\rm i}( S^{c} _{nm})_{;k_{b}} \\
& ( S^{c}_{nm})_{;k_{b}}=\partial_{k_b}S^{c}_{nm}-{\rm i}(\xi^{b}_{nn}-\xi^{b}_{mm})S^{c}_{nm}.
\label{eq:aa3}
\end{aligned}
\end{equation} 
Here, we require that $S^{c}$ is an operator which can be easily handled like the interband of the position operator $\boldsymbol{r}_{e}$.

Based on quantum Liouville equation
\begin{equation}
\begin{aligned}  
i\hbar \frac{d\rho }{dt} =[ H,\rho ]
\label{eq:aa4}
\end{aligned}
\end{equation} 
and discussions above, we can derive the dynamical equation in Bloch basis
\begin{equation}
\begin{aligned} 
\frac{d\rho _{mn}}{dt} +{\rm i}\omega _{mn} \rho _{mn} 
&=-\frac{e}{\hbar }E_{b}(t)( \rho _{mn})_{;k_{b}} \\
&+\frac{{\rm i}e}{\hbar }E_{b}(t)\sum _{l}\left( r_{ml}^{b} \rho _{ln} -\rho _{ml} r_{ln}^{b}\right) \\
&+ \left.\frac{d\rho _{mn}}{dt}\right|_{\text{scat}},
\label{eq:aa5}
\end{aligned}
\end{equation} 
in which $\omega _{mn}=(\varepsilon_{m}-\varepsilon_{n})/\hbar$ and $\varepsilon_{n}$ is the energy of band $n$. The first term in the r.h.s. of equation comes from the commutator of intraband part of position operator and density operator $[r^{b}_{i},\rho]$, the second term comes from the commutator of interband part of position operator and density operator $[r^{b}_{e},\rho]$ and the third term corresponds to the density change due to scattering effects $[H_{\text{\rm scat}},\rho]$. To proceed the discussion, the scattering effects are considered with relaxation time approximation $\left.\frac{d\rho _{mn}}{dt}\right|_{\text{scat}}=-\tau_{s}^{-1}(\rho _{mn}-\rho _{mn}^{(0)})$, where $\tau_{s}$ is the phenomological relaxation time and $\rho _{mn}^{(0)}=\delta_{mn}f_{n}(\varepsilon_{n})$ is the fermi distribution. By iteratively solving this equation, we can access the density correction of each order as $\rho_{mn}=\rho_{mn}^{(0)}+\rho_{mn}^{(1)}+\rho_{mn}^{(2)}+...$. For example, when we insert $\rho_{mn}^{(0)}$ into the second term of the r.h.s. of dynamical equation, we will get a density correction which is linear in electric field and can be marked as $\rho_{mn}^{e}$ since it comes from the interband density correction. $\rho_{mn}^{e}$ is expressed as
\begin{equation}
\begin{aligned} 
\rho_{mn}^{e}=-\frac{e}{\hbar}\frac{f_{nm}r_{mn}^{b}}{ \omega _{mn} -\omega _{\gamma}-{\rm i}\tau_{s}^{-1}} E_{b}^{\gamma}e^{-{\rm i}\omega_{\gamma}t},
\label{eq:aa6}
\end{aligned}
\end{equation} 
where $f_{nm}=f_{n}-f_{m}$ is the difference of Fermi distribution between band $n$ and $m$. If we insert $\rho_{mn}^{e}$ into the first term of the dynamical equation, we will get a second-order density correction $\rho_{mn}^{ei}$. In conclusion, we have $ \rho_{mn}^{(1)}=\rho_{mn}^{i}+\rho_{mn}^{e}$ for linear density correction and $ \rho_{mn}^{(2)}=\rho_{mn}^{ii}+\rho_{mn}^{ie}+\rho_{mn}^{ei}+\rho_{mn}^{ee}$ for second-order density correction. Higher order density corrections can be accessed in a similar way.

The expectation value of electric current is 
\begin{equation}
\begin{aligned} 
j_{a}=-e\sum_{n,m,\boldsymbol{k}}v^{a}_{nm}\rho_{mn},
\label{eq:aa7}
\end{aligned}
\end{equation} 
in which $v^{a}_{nm}$ is the matrix elements of velocity operator. 

Then, the total $n$th-order photocurrent with a frequency $\omega_{n}$ can be expressed as 
\begin{equation}
\begin{aligned} 
& j_a^{(n)}\left(t;\omega_n\right)=j_a^{(n)}\left(\omega_n\right)e^{-{\rm i}{\omega}_{n}t}+j_a^{(n)}\left(-\omega_n\right)e^{{\rm i}{\omega}_{n}t} \\
& j_a^{(n)}\left(\omega_n\right)=\sum_{C\left(\omega_n\right)} \sigma^{a b_1 \ldots b_n}\left(\omega_n ; \omega_{\gamma_1} \cdots \omega_{\gamma_n}\right) E_{b_1}^{\gamma_1} \cdots E_{b_n}^{\gamma_n},
\label{eq:aa8}
\end{aligned}
\end{equation} 
where $\omega_{n}=\omega_{\gamma_1}+\cdots\omega_{\gamma_n}$, and $\sigma^{ab_{1}...b_{n}}(\cdots)$ is the $n$th-order conductivity tensor. The set $C(\omega_n)$ collects all possible selections of $n$ frequencies that sum to $\omega_{n}$. The formula in the second line is the Eq. (\ref{eq:b2}) in the section II.

Below, we take some typical optical responses induced by monochromatic light as the example to show why the general $n$th-order optical conductivity derived from the length-gauge theory can be written in the form of Eq. (\ref{eq:b3}). Linear optical conductivity is 
\begin{equation}
\begin{aligned} 
& j^{\text{opt}}_{a}=\sigma^{ab}_{\text{opt}}(\omega;\omega)E(\omega)e^{-{\rm i}\omega t} \\
& \sigma^{ab}_{\text{opt}}=\frac{{\rm i}e^{2}}{\hbar }\int [ d\boldsymbol{k}]\sum _{n\neq m}f_{nm} \omega _{nm} r_{nm}^{a} r_{mn}^{b}\delta(\omega _{mn} -\omega),
\label{eq:aa9}
\end{aligned}
\end{equation} 
in which $[d\boldsymbol{k}]=d^dk/(2\pi)^d$. This photocurrent comes from the resonant part of $\rho^{e}_{nm}$, which corresponds to the second term of the decomposition $(\omega_{mn}-\omega-\tau_{s}^{-1})^{-1}=\mathcal{P}(\omega_{mn}-\omega)^{-1}+{\rm i}\pi\delta(\omega_{mn}-\omega)$. $\mathcal{P}(...)$ represents the principal part and $\delta(...)$ indicates the Dirac delta function. Hereafter, when we introduce optical responses, the frequency dependence of the optical conductivity in the second line is omitted for brevity.

It is widely acknowledged that shift current and injection current are two contributions that dominate the second-order dc photocurrent when energy of light is larger than the band gap $\hbar\omega>E_{\text{gap}}$ in the clean limit.\cite{sipe2000,watanabe2021} The shift current is 
\begin{equation}
\begin{aligned} 
& j^{\text{shift}}_{a}=\sigma^{abc}_{\text{shift}}(0;\omega,-\omega)E_{b}(\omega)E_{c}(-\omega) \\
& \sigma^{abc}_{\text{shift}}=-\frac{\pi e^{3}}{2\hbar ^{2}}\int [d\boldsymbol{k} ]\sum _{n,m} f_{nm} \bar{R}_{mn}^{a} r_{nm}^{c} r_{mn}^{b} \delta ( \omega _{mn} -\omega ),
\label{eq:aa10}
\end{aligned}
\end{equation} 
in which $\bar{R}_{mn}^{a}=R_{mn}^{a;b} -R_{nm}^{a;c}$ and $R_{mn}^{a;b} =i\partial _{k_{a}}\left(\log r_{mn}^{b}\right) +\xi _{mm}^{a} -\xi _{nn}^{a}$ is the shift vector. Shift current can be viewed as the second-order dc photocurrent that comes from the shift of the real space center of electron, when the electron in the valence band absorbs a photon and jump into condcution band. The injection current is 
\begin{equation}
\begin{aligned} 
& j^{\text{inj}}_{a}=\sigma^{abc}_{\text{inj}}(0;\omega,-\omega)E_{b}(\omega)E_{c}(-\omega)\\
& \sigma_{\text{inj}}^{abc}=-\frac{\tau_{s} \pi e^{3}}{\hbar ^{2}}\int [d\boldsymbol{k} ]\sum _{n,m} f_{nm} \Delta _{mn}^{a} r_{nm}^{c} r_{mn}^{b} \delta ( \omega _{mn} -\omega ),
\label{eq:aa11}
\end{aligned}
\end{equation} 
in which $ \tau_{s}$ is the relaxation time, $\Delta_{nm}^{a} =v_{m}^{a}-v_{n}^{a}$ is the difference of group velocity between band $m$ and $n$. The name injection current comes from the fact that this dc photocurrent grows linearly with illumination time within relaxation time.

Recently, it is recognized that jerk current has important contribuition to the third order dc photocurrent, when there exist a static electric field in addition to the monochromatic light.\cite{fregoso2018} The jerk current can be expressed as
\begin{equation}
\begin{aligned} 
& j^{\text {jerk }}_a=  \sigma_{\text {jerk }}^{a b c d}(0;\omega,-\omega,0) E_b(\omega) E_c(-\omega) E_d(0)\\
& \sigma_{\text {jerk }}^{a b c d}=\frac{2 \pi \tau_{s}^2e^4}{\hbar^3} \int [d\boldsymbol{k} ] \sum_{n, m} f_{n m} \frac{\partial^2 \omega_{n m}}{\partial k^a \partial k^d} r_{n m}^cr_{m n}^b\delta ( \omega _{mn} -\omega ),
\label{eq:aa12}
\end{aligned}
\end{equation} 
which grows quadratically with illumination time within relaxation time. 

Based on the discussions above, it is easy to see that the total $n$th-order photocurrent with a frequency $\omega_{n}$ can be expressed as Eq. (\ref{eq:b2}) and a general $n$th-order optical conductivity can be written in the form of Eq. (\ref{eq:b3}).

\subsection{Symmetry analysis of nonlinear optical conductivity}
\setcounter{equation}{0}
\renewcommand\theequation{B\arabic{equation}} 
Symmetry analysis of nonlinear optical conductivity is straightforward if the symmetry operation is pure spatial. In this case, we can directly perform tensoral transformation on the nonlinear optical conductivity tensor and see the consequence of the symmetry operation. However, things become troublesome when the symmetry operation involves time-reversal because of dissipation.\cite{ahn2020} A proper analysis for such symmetry requires knowledge of a microscopic expression of optical conductivities. For the sake of completeness and clarity, we paraphrase how to analyze the impact of time-reversal of the nonlinear optical conductivity presented in previous literature in this section and illustrate with an example.

A general $n$th-order optical response can be expressed as 
\begin{equation}
\begin{aligned} 
j^{(n)}_{a}(\tilde{\omega}_{n})=\sigma^{ab_{1}...b_{n}}(\tilde{\omega}_{n})E_{b_{1}}(\tilde{\omega}_{\gamma_{1}})\dots E_{b_{n}}(\tilde{\omega}_{\gamma_{n}}).
\label{eq:bb1}
\end{aligned}
\end{equation} 
Here we define $\tilde{\omega}=\omega+{\rm i}\tau_{s}^{-1}$. For brevity, $\sigma^{ab_{1}...b_{n}}(\tilde{\omega}_{n})$ represents $\sigma^{ab_{1}...b_{n}}(\tilde{\omega}_{n};\tilde{\omega}_{\gamma_{1}}\dots\tilde{\omega}_{\gamma_{n}})$. In the discussions below, when we write $\sigma^{ab_{1}...b_{n}}(-\tilde{\omega}_{n})$, it means $\sigma^{ab_{1}...b_{n}}(-\tilde{\omega}_{n};-\tilde{\omega}_{\gamma_{1}}\dots-\tilde{\omega}_{\gamma_{n}})$.

The time-reversal counterpart of Eq.(\ref{eq:bb1}) is
\begin{equation}
\begin{aligned} 
j^{{\prime},(n)}_{a}(\tilde{\omega}_{n})=\sigma^{{\prime},ab_{1}...b_{n}}(\tilde{\omega}_{n})E_{b_{1}}^{\prime}(\tilde{\omega}_{\gamma_{1}})\dots E_{b_{n}}^{\prime}(\tilde{\omega}_{\gamma_{n}}).
\label{eq:bb2}
\end{aligned}
\end{equation} 

Since time-reversal operation $\Theta$ requires
\begin{equation}
\begin{aligned} 
& j_{a}(\tilde{\omega}_{n}){\rightarrow}j^{\prime}_{a}(\tilde{\omega}_{n})=-j_{a}(-\tilde{\omega}_{n}) \\
& E_{a}(\tilde{\omega}_{\gamma_{i}}){\rightarrow}E^{\prime}_{a}(\tilde{\omega}_{\gamma_{i}})=E_{a}(-\tilde{\omega}_{\gamma_{i}}),
\label{eq:bb3}
\end{aligned}
\end{equation} 
Eq.(\ref{eq:bb2}) can be further written as 
\begin{equation}
\begin{aligned} 
    -j^{(n)}_{a}(-\tilde{\omega}_{n})=\sigma^{{\prime},ab_{1}...b_{n}}(\tilde{\omega}_{n})E_{b_{1}}(-\tilde{\omega}_{\gamma_{1}})\dots E_{b_{n}}(-\tilde{\omega}_{\gamma_{n}}).
\label{eq:bb4}
\end{aligned}
\end{equation} 

On the other hand, we can rewrite Eq.(\ref{eq:bb1}) as 
\begin{equation}
\begin{aligned} 
j^{(n)}_{a}(-\tilde{\omega}_{n})=\sigma^{ab_{1}...b_{n}}(-\tilde{\omega}_{n})E_{b_{1}}(-\tilde{\omega}_{\gamma_{1}})\dots E_{b_{n}}(-\tilde{\omega}_{\gamma_{n}}).
\label{eq:bb5}
\end{aligned}
\end{equation} 
Thus, it can be concluded that the time-reversal operation requires the optical conductivity to transform as follows
\begin{equation}
\begin{aligned} 
\sigma^{{\prime},ab_{1}...b_{n}}(\tilde{\omega}_{n})=-\sigma^{ab_{1}...b_{n}}(-\tilde{\omega}_{n}).
\label{eq:bb6}
\end{aligned}
\end{equation} 

Let's take linear injection current (the part of injection current that corresponds to the $g_{nm}^{cb}$ in the decomposition of $r_{nm}^{c} r_{mn}^{b}$) as the example to illustrate how to analyze optical conductivity under time-reversal operation. Linear injection current is
\begin{equation}
\begin{aligned} 
\sigma_{\text{inj,L}}^{abc}=\frac{\tau_{s} e^{3}}{\hbar ^{2}}\int [d\boldsymbol{k} ]\sum _{n,m} f_{nm} \Delta _{mn}^{a} g_{nm}^{cb} \frac{\tau_{s}^{-1}}{(\omega_{mn}-\omega)^2+\tau_{s}^{-2}}.
\label{eq:bb7}
\end{aligned}
\end{equation} 
Here, we express the Dirac delta function as
\begin{equation}
\begin{aligned} 
\delta(\omega_{mn}-\omega)=\frac{1}{\pi}\frac{\tau_{s}^{-1}}{(\omega_{mn}-\omega)^2+\tau_{s}^{-2}}
\label{eq:bb8}
\end{aligned}
\end{equation} 
for the convenience of analyzing $\tau_{s}{\rightarrow}-\tau_{s}$.

According to Eq.(\ref{eq:bb6}), for the system illuminated by monochromatic light, the second-order dc optical conductivity under the time-reversal operation is
\begin{equation}
\begin{aligned} 
\sigma^{{\prime},abc}(\tilde{\omega};\tilde{\omega}_{1},\tilde{\omega}_{2})=-\sigma^{abc}(-\tilde{\omega};-\tilde{\omega}_{1},-\tilde{\omega}_{2}),
\label{eq:bb9}
\end{aligned}
\end{equation} 
in which $\tilde{\omega}=\tilde{\omega}_{1}+\tilde{\omega}_{2}$, $\tilde{\omega}_{1}=\omega+{\rm i}\tau_{s}^{-1}$, $\tilde{\omega}_{2}=-\omega+{\rm i}\tau_{s}^{-1}$. Then we have
\begin{equation}
\begin{aligned} 
&\sigma_{\text{inj,L}}^{\prime,abc} \\
&=-\frac{(-\tau_{s}) e^{3}}{\hbar ^{2}}\int [d\boldsymbol{k} ]\sum _{n,m} f_{nm} \Delta _{mn}^{a} g_{nm}^{cb} \frac{(-\tau_{s})^{-1}}{(\omega_{mn}+\omega)^2+(-\tau_{s})^{-2}} \\
&=-\frac{\tau_{s} e^{3}}{\hbar ^{2}}\int [d\boldsymbol{k} ]\sum _{n,m} f_{nm} \Delta _{mn}^{a} g_{nm}^{cb} \frac{\tau_{s}^{-1}}{(\omega_{mn}-\omega)^2+\tau_{s}^{-2}} \\
&=-\sigma_{\text{inj,L}}^{abc}.
\label{eq:bb10}
\end{aligned}
\end{equation} 
In the derivation above, the second equality takes into account the exchange of the $m$ and $n$ indices, as well as the symmetric property of $g_{nm}^{cb}$ under the exchange of these indices. Through this analysis, we have verified that the linear injection current changes sign under time-reversal operation.  Then we further find that linear injection current is symmetric under space-time inversion $P\Theta$, which indicates it corresponds to $\sigma_{0}$. Other optical conductivities can be analyzed in a similar way. Here we emphasize that, though linear injection current is identified as $\sigma_{0}$,  linear injection current is identified as $\sigma_{\chi}$. The correspondence between $g_{nm}^{cb}/\Omega_{nm}^{cb}$ and $\sigma_{0}/\sigma_{\chi}$ needs examination of the microscopic expression of the conductivities.

\subsection{Extend the discussions of Eqs. (\ref{eq:b5}-\ref{eq:b6}) and Table \ref{tab:1} to the multi-Weyl nodes}
\setcounter{equation}{0}
\renewcommand\theequation{C\arabic{equation}} 
In addition to the normal Weyl nodes described by Eq. (\ref{eq:b1}) with chiral charges $\chi=\pm1$, there also exist multi-Weyl nodes that have larger chiral charges with $\chi=\pm2,\pm3$, which are protected by rotation $C_{4}$ or $C_{6}$.\cite{fang2012} A multi-Weyl node is generally described by\cite{li2021} 
\begin{equation}
\begin{aligned}  
\mathcal{H}_n=t k_z+w k_{\|}^2+v_z k_z \sigma_z + v(k_{+}^n \sigma_{+} + k_{-}^n \sigma_{-}).
\label{eq:cc1}
\end{aligned}
\end{equation}
The eigenstates of the multi-Weyl node are $\varepsilon_{s}=t k_z + w k_{\|}^2 + s\sqrt{v_z^2 k_z^2+v^2 k_{\|}^{2 n}}$, where $s = \pm1$ corresponds to conduction and valence bands, respectively. $k_{\|}=\sqrt{k_x^2 + k_y^2}$ and $k_{\pm}=k_x \pm ik_y$. The subscript $n = 2, 3$ denotes the double and triple Weyl nodes, respectively. The first two terms refer to the linear and quadratic tilting terms. Contrast to the linear tilting term, the quadratic tilting term with $w>0$ ($w<0$) always pushes up (down) spectrum for any $k$-plane with a given $k_z$. When $|t| > |v_z|$/$|w| > |v|$, the spectrum is overtilted, leading to a type-II/type-III Weyl node.

If we want to discuss how the tilt and chirality of the multi-Weyl node affect its optical conductivity, we need to identify the definitions of the reversal of tilt and chirality for the multi-Weyl node. According to the previous discussions, for a normal Weyl node described by Eq. (\ref{eq:b1}), the chirality is reversed while the spectrum is unchanged under the space-time inversion $P\Theta$ and tilt is reversed while chirality remains unaffected under the time-reversal $\Theta$. Since Eqs. (\ref{eq:b5}-\ref{eq:b6}) and Table \ref{tab:1} are equivalent to the symmetry analysis, the discussions can be generalized to the multi-Weyl node by examining how they transformed under $P\Theta$ and $\Theta$. By doing so, chirality reversal for the multi-Weyl node can be identified as $(v_z,v)\mapsto(-v_z,(-1)^{n}v)$, and tilt reversal as $(t,w)\mapsto(-t,w)$.

\subsection{Details of calculating $\tilde{\sigma}_{0}$ and $\tilde{\sigma}_{\chi}$ of Weyl nodes with Landau levels}
\setcounter{equation}{0}
\renewcommand\theequation{D\arabic{equation}} 

Introducing the $B$-field in $z$-direction with vector potential $\bm a = Bx \hat y/\hbar$, the Weyl nodes in ferromagnetic $\rm MnBi_{2}Te_{4}$ can be described by Eq. (\ref{eq:c3}). Then the energy of the $n$th ($n$ an integer) Landau level at $\vec k =[k_y,k_z]$ is
\begin{equation}
\varepsilon_{n\vec k} = t v_t k_z +\chi \operatorname{sgn} (n)\eta_n,
\label{eq:s4}
\end{equation}
in which $\eta_n = \sqrt{2|n| \hbar^2 v_\perp^2/\ell^2+\eta_{\parallel}^2}\ (n\neq0)$, $\ell = \sqrt{\hbar/eB}$ is the magnetic length, $\eta_{0}=\eta_{\parallel}= v_{\parallel}k_z$, and the sign function $\operatorname{sgn}(n)$ equals $-1$ if $n=0$. The corresponding wavefunction is
\begin{equation}
\Psi_{n\vec k}(x) = \frac{e^{\mathrm i (k_yy+k_zz)}}{\sqrt{L_yL_z}}\begin{bmatrix}
    \operatorname{sgn}(n) u_n \psi_{|n|-1}(x') \\
    \text{i} w_n \psi_{|n|}(x')
\end{bmatrix},
\label{eq:s5}
\end{equation}
where $x'=x/\ell+k_y\ell$, $\sqrt{\ell}\psi_n(x)$ is the Hermite-Gaussian function, and the coefficients $u/ w_n=\sqrt{\frac{1}{2}\left(1\pm \operatorname{sgn}(n)\frac{\eta_{\parallel}}{\eta_n}\right)}$. The velocity matrix elements that essential for nonlinear optical conductivities are
\begin{equation}
\begin{aligned}
    v_{x,mn \vec k} &= \mathrm i \chi \tilde{v}_\perp
    \left[\delta_{|m|,|n|+1} \operatorname{sgn}(m)u_mw_n-(m\leftrightarrow n)\right], \\
    v_{y,mn \vec k} &= \chi \tilde{v}_\perp
    \left[\delta_{|m|,|n|+1} \operatorname{sgn}(m)u_mw_n+(m\leftrightarrow n)\right], \\
v_{z,mn \vec k} &= \delta _{|m|,|n|}\left(( \tilde{v}_{t} +\chi \tilde{v}_{\parallel })\operatorname{sgn} (m)\operatorname{sgn} (n)u_{m} u_{n} \right. \\
&+ \left. ( \tilde{v}_{t} -\chi \tilde{v}_{\parallel }) w_{n} w_{m}\right)
\label{eq:s6}
\end{aligned}
\end{equation}
 with $\tilde{v}=v/\hbar$. From Eq. (\ref{eq:s6}), we can see that the optical transitions are nonzero only between Landau levels that satisfy $|m|-|n|=\pm 1$.

\bibliographystyle{statto}
% \bibliography{./ref}
\end{document}